\documentclass[12pt]{article}

\usepackage{amsmath}
\usepackage{amssymb}
\usepackage{bm}
\usepackage{caption}
\usepackage{color}
\usepackage{enumerate}
\usepackage{graphicx}
\usepackage{natbib}
\usepackage{subcaption}
\usepackage{url}
\usepackage{multirow}
\usepackage{comment}

\makeatletter
\let\save@mathaccent\mathaccent
\newcommand*\if@single[3]{%
  \setbox0\hbox{${\mathaccent"0362{#1}}^H$}%
  \setbox2\hbox{${\mathaccent"0362{\kern0pt#1}}^H$}%
  \ifdim\ht0=\ht2 #3\else #2\fi
  }
\newcommand*\rel@kern[1]{\kern#1\dimexpr\macc@kerna}
\newcommand*\widebar[1]{\@ifnextchar^{{\wide@bar{#1}{0}}}{\wide@bar{#1}{1}}}
\newcommand*\wide@bar[2]{\if@single{#1}{\wide@bar@{#1}{#2}{1}}{\wide@bar@{#1}{#2}{2}}}
\newcommand*\wide@bar@[3]{%
  \begingroup
  \def\mathaccent##1##2{%
    \let\mathaccent\save@mathaccent
    \if#32 \let\macc@nucleus\first@char \fi
    \setbox\z@\hbox{$\macc@style{\macc@nucleus}_{}$}%
    \setbox\tw@\hbox{$\macc@style{\macc@nucleus}{}_{}$}%
    \dimen@\wd\tw@
    \advance\dimen@-\wd\z@
    \divide\dimen@ 3
    \@tempdima\wd\tw@
    \advance\@tempdima-\scriptspace
    \divide\@tempdima 10
    \advance\dimen@-\@tempdima
    \ifdim\dimen@>\z@ \dimen@0pt\fi
    \rel@kern{0.6}\kern-\dimen@
    \if#31
      \overline{\rel@kern{-0.6}\kern\dimen@\macc@nucleus\rel@kern{0.4}\kern\dimen@}%
      \advance\dimen@0.4\dimexpr\macc@kerna
      \let\final@kern#2%
      \ifdim\dimen@<\z@ \let\final@kern1\fi
      \if\final@kern1 \kern-\dimen@\fi
    \else
      \overline{\rel@kern{-0.6}\kern\dimen@#1}%
    \fi
  }%
  \macc@depth\@ne
  \let\math@bgroup\@empty \let\math@egroup\macc@set@skewchar
  \mathsurround\z@ \frozen@everymath{\mathgroup\macc@group\relax}%
  \macc@set@skewchar\relax
  \let\mathaccentV\macc@nested@a
  \if#31
    \macc@nested@a\relax111{#1}%
  \else
    \def\gobble@till@marker##1\endmarker{}%
    \futurelet\first@char\gobble@till@marker#1\endmarker
    \ifcat\noexpand\first@char A\else
      \def\first@char{}%
    \fi
    \macc@nested@a\relax111{\first@char}%
  \fi
  \endgroup
}
\makeatother
\bibpunct{(}{)}{;}{a}{,}{,}

\renewcommand{\baselinestretch}{1.6}

\newtheorem{theorem}{Theorem}
\newtheorem{lemma}{Lemma}
\newtheorem{proposition}{Proposition}
\def\beq{\begin{equation}}
\def\eeq{\end{equation}}
\def\beqr{\begin{eqnarray}}
\def\eeqr{\end{eqnarray}}
\def\beqrs{\begin{eqnarray*}}
\def\eeqrs{\end{eqnarray*}}
\def\bet{\begin{theorem}}
\def\eet{\end{theorem}}
\def\bel{\begin{lemma}}
\def\eel{\end{lemma}}
\def\bep{\begin{proposition}}
\def\eep{\end{proposition}}
\def\bg{\begin{figure}[tbph]\begin{center}}
\def\eg{\end{center}\end{figure}}

\def\bc{\begin{center}}
\def\ec{\end{center}}

\def\A1{A{\mathbf 1}}
\def\1{{\mathbf 1}}

\renewcommand{\baselinestretch}{2}

\newcommand{\bX}{\mathbf{X}}
\newcommand{\qed}{\hfill \mbox{\raggedright \rule{.07in}{.1in}}}
\newcommand{\SMD}{\mbox{\tiny{SMD}}}
\newcommand{\PSW}{\mbox{\tiny{PSW}}}

\DeclareMathOperator\bE{\mathbb E} 

\makeatletter
\newcommand{\figcaption}{\def\@captype{figure}\caption}
\newcommand{\tabcaption}{\def\@captype{table}\caption}
\makeatother

\newcommand{\one}{\mathbf{1}}

\begin{document}

\begin{center}
\vspace*{-2.5cm}

{\Large Subgroup Balancing Propensity Score}

\medskip

Jing Dong \\
{\it Industrial and Commercial Bank of China Ltd. }\\
Junni L. Zhang\\
{\it Guanghua School of Management and Center for Statistical Science, Peking University }\\
Fan Li\footnotetext[2]{Jing Dong (email: dongjing@icbc.com.cn) is Post-doctor at Industrial and Commercial Bank of China, Beijing 100040, P. R. China.  Junni L. Zhang (email: zjn@gsm.pku.edu.cn) is Associate Professor at Guanghua School of Management and Center for Statistical Science, Peking University, Beijing 100871, P. R. China.  Fan Li (email: fl35@stat.duke.edu) is Associate Professor at Department of Statistical Science, Duke University, USA.  The authors are grateful to Andrea Mercatanti for providing the SHIW data. Part of this paper was written when Jing Dong was an exchange student at Duke University under the High-Level Graduate Student Scholarship of China Scholarship Council.
}\\
{\it Department of Statistical Science, Duke University}{}

\end{center}

\date{}

{\centerline{ABSTRACT}
\noindent We investigate the estimation of subgroup treatment effects with observational data. Existing propensity score matching and weighting methods are mostly developed for estimating overall treatment effect. Although the true propensity score should balance covariates for the subgroup populations, the estimated propensity score may not balance covariates for the subgroup samples. We propose the subgroup balancing propensity score (SBPS) method, which selects, for each subgroup, to use either the overall sample or the subgroup sample to estimate propensity scores for units within that subgroup, in order to optimize a criterion accounting for a set of covariate-balancing conditions for both the overall sample and the subgroup samples. We develop a stochastic search algorithm for the estimation of SBPS when the number of subgroups is large. We demonstrate through simulations that the SBPS can improve the performance of propensity score matching in estimating subgroup treatment effects. We then apply the SBPS method to data from the Italy Survey of Household Income and Wealth (SHIW) to estimate the treatment effects of having debit card on household consumption for different income groups.

\vspace*{0.3cm}

}

\clearpage

\section{Introduction}





Estimating the causal effects of a treatment or intervention is a central goal in many observational studies. Most existing research focuses on treatment effects for the overall population, such as the average treatment effect (ATE) and the average treatment effect for the treated (ATT). However, heterogeneity in treatment effects across different subpopulations is common and accurate estimation of subgroup effects may provide more refined picture of how the treatment works and facilitate better design of relevant programs in the future \citep{barnard2003principal,jin2010modified}. For example, in our application, we estimate the effects of possessing debit cards on household consumption, which vary significantly across different income groups. In this paper, we focus on subpopulations that are explicitly defined by observed covariates rather than latent variables such as compliance status \citep{ImbensAngrist94}.


The propensity score is the probability of receiving the treatment given the covariates. \cite{rosenbaum1983central} proved that the propensity score has two important properties: (i) it is a balancing score (in fact, the coarsest balancing score), that is, balancing the distribution of the propensity score between the treatment groups would balance the distribution of all covariates; and (ii) if the treatment assignment is unconfounded given the covariates, then it is also unconfounded given the propensity score. Therefore, the propensity score can be viewed as a summary score of the multivariate covariates.  These properties ensure that matching or weighting based on the propensity score effectively mimics a randomized experiment, and thus the treatment effects can be estimated by comparing observed outcomes for the treated and control groups adjusting for propensity scores instead of the whole set of covariates. 

Theoretically, we can show that in addition to balancing covariates for the overall population, the propensity score also balances covariates for the subgroup populations (details in Section \ref{sec:ps}), hence after propensity score matching or weighting, we can compare the treated and control units within each subgroup to draw inference about treatment effects for that subgroup. However, in practice, the propensity scores must be estimated from the finite sample. Traditionally, it has been estimated using the overall sample. The estimated propensity score may balance covariates for the overall sample, but may not balance covariates for the subgroup samples. On the other hand, if the propensity scores for units in a subgroup are estimated using the subgroup sample, covariates may be better balanced for this subgroup sample, but such estimation may have larger variability since the sample size for the subgroup sample is smaller than that for the overall sample.

In order to better estimate the propensity score using finite samples, we propose the subgroup balancing propensity score (SBPS) method, a hybrid approach combining propensity score estimation from the overall sample and the subgroup samples. For each subgroup, we select to use either the overall sample or the subgroup sample to estimate propensity scores for units within this subgroup. The combination of estimation samples is chosen by optimizing a criterion that accounts for a set of covariate-balancing moment conditions for both the overall sample and the subgroup samples. We develop a stochastic search algorithm for the estimation of SBPS when the number of subgroups is large.  Covariate-balancing moment conditions have been previously used in the entropy balancing method \citep{hainmueller2011entropy} and the covariate balancing propensity score (CBPS) method \citep{imai2014covariate} in the context of estimating overall (instead of subgroup) treatment effect estimands. 

The rest of this paper is organized as follows. Section 2 defines the subgroup ATT estimands and present the estimators based on propensity score matching or weighting. Section 3 presents the SBPS method and the estimation algorithm. Section 4 conducts a simulation study to compare the performance of the SBPS method with the traditional method and the CBPS method. In Section 5, we apply the SBPS method to the Italy Survey of Household Income and Wealth (SHIW) data to evaluate the effects of debit card possession on monthly household consumption for different income groups. We also conduct a simulation based on the SHIW data to gain more insights of the performance of the proposed method. Section 6 concludes.

\section{Estimating Subgroup Causal Effects}\label{sec:ps}


%


%

Consider a sample of $N$ units, consisting of $N_t$ treated and $N_c$ control units. Suppose that the population can be partitioned into subpopulations based on certain covariates and the interest is to estimate the treatment effects for these subpopulations. Specifically, in this paper we focus on subpopulations defined by a single covariate $U$ with $R$ number of possible values $u\in \{u_1,...,u_R\}$. For each unit $i$, denote $G_i$ the label of its subgroup, with $G_i=r$ if $U_i=u_r$. Also denote $Z_i$ the treatment indicator, $\bm{X}_i=(X_{i1},\cdots,X_{iK})'$ the covariates excluding $U$, and $Y_i(z)$ the potential outcome corresponding to treatment $z$ for $z\in \{0,1\}$. For each unit, only the potential outcome under the assigned treatment, $Y_i=Y_i(Z_i)$, is observed and the other potential outcome is missing. Let $N_r$ denote the number of units in subgroup $r$, with $N_{r,t}$ treated units and $N_{r,c}$ control units for $r=1,\cdots,R$.
We are interested in estimating the subgroup ATTs:
\begin{equation}\label{eq:ATT-subgroup-defn}
\tau_{r}=\bE\left[Y(1)-Y(0)|G=r,Z=1\right],\quad r=1,\cdots,R.
\end{equation}

The propensity score is the probability of receiving the treatment given the covariates, which includes $\bm{X}$ and the subgroup label $G$,
\begin{equation}\label{eq:ps}
e(\bm{X},G)=\Pr(Z=1|\bm{X},G).
\end{equation}
Applying results from \cite{rosenbaum1983central}, the above propensity score is a balancing score in that the covariates and the subgroup label are independent of the treatment variable conditional on the propensity score:
\begin{equation}\label{eq:bal-property}
Z \perp \{\bm{X},G\} |e(\bm{X},G).
\end{equation}
Assume that there is overlap in the propensity score distribution between the treated and control groups, that is, $0<e(\bm{X},G)<1$ for any $\bm{X}$ and $G$. Assume also that the treatment assignment is unconfounded given the covariates:
\begin{equation}\label{eq:unconfounded-assignment}
Z \perp \{Y(1),Y(0)\} |\{\bm{X},G\}.
\end{equation}
Then we have the following result regarding propensity scores for subgroups.


\textbf{Result 1}.  The propensity score satisfies:
\begin{equation}\label{eq:bal-property2}
\bm{X}\perp Z|\{G=r,e(\bm{X},G)\} \quad \mbox{for}\ \ r=1,\cdots,R.
\end{equation}
\textbf{Proof.} Equation \eqref{eq:bal-property} is equivalent to $f(\bm{X},G|e(\bm{X},G),Z=1)=f(\bm{X},G|e(\bm{X},G),Z=0)$. Marginalizing out $\bX$ gives  $f(G|e(\bm{X},G),Z=1)=f(G|e(\bm{X},G),Z=0)$, and therefore we have
\begin{equation*}
f(\bm{X}|G=r,e(\bm{X},G),Z=1)=f(\bm{X}|G=r,e(\bm{X},G),Z=0)\ \ \mbox{for}\ \ r=1,\cdots,R,
\end{equation*}
which is equivalent to equation \eqref{eq:bal-property2}. $\qed$\\
Result 1 implies that the propensity score not only balances the distribution of $\bm{X}$ and $G$ for the overall population, but also balances the distribution of $\bm{X}$ in each subgroup.

We consider propensity score matching within each subgroup, that is, for each treated unit, select a matching control unit from the same subgroup based on the propensity score. Therefore, within each subgroup in the matched population, the distributions of the propensity score are the same between the treated and control groups, that is, for $r=1,\cdots,R$,
\begin{equation}\label{eq:matching-balanceps}
f(e(\bm{X},G)|G=r,Z=1)=f(e(\bm{X},G)|G=r,Z=0).
\end{equation}
Combining \eqref{eq:matching-balanceps} and \eqref{eq:bal-property2}, for $r=1,\cdots,R$, we have
\begin{equation}\label{eq:matching-balanceX}
\begin{aligned}
&f(\bm{X}|G=r,Z=1)=\int f(\bm{X}|G=r,e(\bm{X},G))f(e(\bm{X},G)|G=r,Z=1)d e(\bm{X},G)\\
&=\int f(\bm{X}|G=r,e(\bm{X},G))f(e(\bm{X},G)|G=r,Z=0)d e(\bm{X},G)=f(\bm{X}|G=r,Z=0).
\end{aligned}
\end{equation}
Hence the distribution of $\bm{X}$ is balanced within each subgroup in the matched population. Since the distribution of $G$ is automatically balanced in the overall matched population, the distribution of $\bm{X}$ is also balanced in the overall matched population. It is straightforward to show that
\begin{equation}\label{eq:ATT-subgroup-est}
\tau_r=\bE^{(m)}\left[Y|G=r,Z=1\right]-
\bE^{(m)}\left[Y|G=r,Z=0\right],
\end{equation}
where $\bE^{(m)}$ indicates taking expectation over the matched population.

We also consider propensity score weighting in which each treated unit is weighted by one and each control unit is weighted by $w(\bm{X},G)=e(\bm{X},G)/(1-e(\bm{X},G))$. It is easy to show that the following set of moment conditions hold:
\begin{equation}
\begin{aligned}\label{eq:moment-weighting}
M_k^{\PSW}&\equiv \bE\left[ZX_k-(1-Z)\frac{e(\bm{X},G)}{1-e(\bm{X},G)}X_k\right]=0,\quad k=1,\cdots,K;\\
M_{(r)}^{\PSW}&\equiv \bE\left[Z\one\{G=r\}-(1-Z)\frac{e(\bm{X},G)}{1-e(\bm{X},G)}\one\{G=r\}\right]=0,\quad r=1,\cdots,R;\\
M_{r,k}^{\PSW}&\equiv \bE\left[\one\{G=r\}\left(ZX_k-(1-Z)\frac{e(\bm{X},G)}{1-e(\bm{X},G)}X_k\right)\right]=0,\quad r=1,\cdots,R,\ k=1,\cdots,K.
\end{aligned}
\end{equation}
Therefore, the specified weights balance $\bm{X}$ for the overall population (reflected by $M^{\PSW}_k=0$), balance $G$ for the overall population (reflected by $M^{\PSW}_{(r)}=0$), and balance $\bm{X}$ for the subgroup populations (which is reflected by $M_{r,k}^{\PSW}=0$).

\textbf{Result 2}.  The subgroup ATTs can be written as
\begin{equation}\label{eq:psw-estimand}
\tau_r=\frac{\bE\left[ZY|G=r\right]}{\text{Pr}(Z=1|G=r)}-\frac{\bE\left[(1-Z)\frac{e(\bm{X},G)}{1-e(\bm{X},G)}Y\Big|G=r\right]}{\text{Pr}(Z=1|G=r)},
\end{equation}
where the denominator can also be written as
\begin{equation}\label{eq:psw-probtreated}
\text{Pr}(Z=1|G=r)=\bE\left[(1-Z)\frac{e(\bm{X},G)}{1-e(\bm{X},G)}\Big|G=r\right].
\end{equation}
\textbf{Proof.} We have
\begin{equation}
\begin{aligned}\label{eq:Result2-eq1}
&\bE\left[ZY-(1-Z)\frac{e(\bm{X},G)}{1-e(\bm{X},G)}Y\Big|G=r\right]\\
=& \bE\left[ZY(1)-(1-Z)\frac{e(\bm{X},G)}{1-e(\bm{X},G)}Y(0)\Big|G=r\right]\\
\end{aligned}
\end{equation}
Because $e(\bm{X},G)=\bE[Z|\bm{X},G]$ and equation \eqref{eq:unconfounded-assignment} holds, we can easily show that \eqref{eq:Result2-eq1}
is equivalent to
\begin{equation}
\begin{aligned}
& \bE\left\{e(\bX,G)\bE\left[Y(1)-Y(0)\big|\bm{X},G=r\right]\Big|G=r\right\}\\
=&\bE\left\{\bE\left[Z(Y(1)-Y(0))\big|\bm{X},G=r\right]\Big|G=r\right\}\\
=&\bE\left\{Z(Y(1)-Y(0))\Big|G=r\right\}\\
=&\text{Pr}(Z=1|G=r)\bE\left[Y(1)-Y(0)|G=r,Z=1\right].
\end{aligned}
\end{equation}
Hence we have \eqref{eq:psw-estimand}.

For the second part, we can easily show that
\begin{equation}
\begin{aligned}
&\bE\left[(1-Z)\frac{e(\bm{X},G)}{1-e(\bm{X},G)}\Big|G=r\right]
=&\bE\left[e(\bm{X},G)|G=r\right].
\end{aligned}
\end{equation}
Because $e(\bm{X},G)=\bE[Z|\bm{X},G]$ and $\bE[Z|G=r]=\text{Pr}(Z=1|G=r)$, we have \eqref{eq:psw-probtreated}.$\qed$

The above discussions are in the setting of true propensity scores and population distributions. In observational studies, the propensity scores  must be estimated from a finite sample.  Let $\hat{e}_i$ denote the estimated propensity score for unit $i$. When propensity score matching within each subgroup is conducted, $\tau_r$ can be estimated by direct comparison:
\begin{equation}\label{eq:est-matching}
\hat{\tau}_r^{Dir}=\widebar{Y}^{(m)}_{r,t}-\widebar{Y}^{(m)}_{r,c},
\end{equation}
where $\widebar{Y}^{(m)}_{r,t}$ and $\widebar{Y}^{(m)}_{r,c}$ denote the mean observed outcomes for the treated and control units in the matched sample for subgroup r. When propensity score weighting is conducted, $\tau_r$ can be estimated by:
\begin{equation}\label{eq:HT-estimator}
\hat{\tau}_r^{PSW}=\widebar{Y}_{r,t}-\frac{\sum_{G_i=r,Z_i=0} \frac{\hat{e}_i}{1-\hat{e}_i}Y_i}{\sum_{G_i=r,Z_i=0} \frac{\hat{e}_i}{1-\hat{e}_i}},
\end{equation}
where $\widebar{Y}_{r,t}$ is the mean observed outcome for the treated units in the sample for subgroup $r$, which is used to estimate the first term in \eqref{eq:psw-estimand}, and the second term in \eqref{eq:HT-estimator} is used to estimate the second term in \eqref{eq:psw-estimand}.

\section{Subgroup Balancing Propensity Score}


The propensity score is commonly estimated by fitting a logistic regression model using the overall sample,
\begin{equation}\label{eq:psmodel1}
\mbox{logit}\left[e(\bm{X},G)\right]
=\sum_{r=1}^R \delta_{r}\one\{G=r\} +\bm{\alpha}^{\top}\bm{X}.
\end{equation}
But the estimated propensity scores often do not completely balance covariates in the overall sample, and the imbalance can be particularly severe in subgroup samples. Consequently, estimates of subgroup effects based on these estimated propensity scores may be biased. An alternative approach is to estimate the propensity score separately within each subgroup, for example, by fitting a logistic model to each subgroup sample:
\begin{equation}\label{eq:psmodel2}
\mbox{logit}\left[e(\bm{X},G)\right]
=\delta_{r} +\bm{\alpha}_r^{\top}\bm{X}, \quad r=1,...,R.
\end{equation}
Matching or weighting based on the subgroup-fitted propensity scores is expected to lead to better balance of covariates within each subgroup and thus smaller biases in causal estimates. However, due to the smaller sample sizes of the subgroups, the ensuing causal estimates may have larger variances.

To address this bias-variance tradeoff, we propose the SBPS as a hybrid approach to adaptively choose between the overall sample fit and the subgroup sample fit by optimizing a criterion that accounts for covariate balance for both the overall sample and the subgroup samples. Below we introduce two criteria, one based on matching and one based on weighting.


\subsection{Matching-based Criterion}
When propensity score matching within each subgroup is conducted, treatment assignment can be treated as random for the overall matched population and for the matched population for each subgroup $r$. Hence we have the following set of moment conditions based on the standardized mean difference (SMD) measure \citep{Rosenbaum&Rubin1985}:
\begin{equation}
\begin{aligned}\label{eq:moment-matching}
M^{\SMD}_k&\equiv \bE^{(m)}\left[\frac{ZX_k-(1-Z)X_k}{\sigma_{k,t}}\right]=0,\quad k=1,\cdots,K;\\
M^{\SMD}_{r,k}&\equiv \bE^{(m)}\left\{\frac{\one\{G=r\}\left[ZX_k-(1-Z)X_k\right]}{\sigma_{r,k,t}}\right\}=0,\quad r=1,\cdots,R,\ k=1,\cdots,K,
\end{aligned}
\end{equation}
where $\sigma_{k,t}$ and $\sigma_{r,k,t}$ denote the standard deviation of $X_k$ for the treated units in the overall population and in the population for subgroup $r$, respectively. The condition $M^{\SMD}_k=0$ reflects balancing of $X_k$ for the overall matched population, and the condition $M^{\SMD}_{r,k}=0$ reflects balancing of $X_k$ for the matched population for subgroup $r$.

For unit $i$ in the matched sample, let $Z_i^{(m)}$ and $G_i^{(m)}$ denote the treatment indicator and the subgroup label, let $x_{ik}^{(m)}$ denote the value of the $X_k$, and let $w_{i}^{(m)}$ denote the matching weight, with $w_{i}^{(m)}=1$ for treated units. Denote $N_t^{(m)}$ and  $N_{r,t}^{(m)}$ the number of treated units in the overall matched sample and in the matched sample for subgroup $r$, respectively.  We have $\sum_{Z_i^{(m)}=0} w_{i}^{(m)}=N_t^{(m)}$ and $\sum_{G_i^{(m)}=r,Z_i^{(m)}=0} w_{i}^{(m)}=N_{r,t}^{(m)}$.
Let $\widebar{x}^{(m)}_{k,t}=\sum_{Z_i^{(m)}=1}x_{ik}^{(m)}/N_t^{(m)}$ and $\widebar{x}^{(m)}_{k,c}=\sum_{Z_i^{(m)}=0}w_i^{(m)}x_{ik}^{(m)}/N_t^{(m)}$ denote the means of $x_{ik}^{(m)}$ for the treated and weighted control units in the overall matched sample, respectively, and let $\widebar{x}^{(m)}_{r,k,t}=\sum_{G_i^{(m)}=r,Z_i^{(m)}=1}x_{ik}^{(m)}/N_{r,t}^{(m)}$ and $\widebar{x}^{(m)}_{r,k,c}=\sum_{G_i^{(m)}=r,Z_i^{(m)}=0}w_i^{(m)}x_{ik}^{(m)}/N_{r,t}^{(m)}$ denote the means of $x_{ik}^{(m)}$ for the treated and weighted control units in the matched sample for subgroup $r$, respectively. Let $\widehat{\sigma}_{k,t}$ and $\widehat{\sigma}_{r,k,t}$ denote the sample standard deviation of $X_k$ for the treated units in the overall sample and in the sample for subgroup $r$, respectively. The estimates of the moments in \eqref{eq:moment-matching} are
\begin{equation*}
\begin{aligned}
\widehat{M}^{\SMD}_k&=\frac{1}{2N_t^{(m)}}\left[\frac{N_t^{(m)}\widebar{x}^{(m)}_{k,t}-N_t^{(m)}\widebar{x}^{(m)}_{k,c}}{\widehat{\sigma}_{k,t}}\right]=\frac{1}{2}\left[\frac{\widebar{x}^{(m)}_{k,t}-\widebar{x}^{(m)}_{k,c}}{\widehat{\sigma}_{k,t}}\right],\\
\widehat{M}^{\SMD}_{r,k}&=\frac{1}{2N_{t}^{(m)}}\left[\frac{N_{r,t}^{(m)}\widebar{x}^{(m)}_{r,k,t}-N_{r,t}^{(m)}\widebar{x}^{(m)}_{r,k,c}}{\widehat{\sigma}_{r,k,t}}\right]=\frac{1}{2}\frac{N_{r,t}^{(m)}}{N_t^{(m)}}\left[\frac{\widebar{x}^{(m)}_{r,k,t}-\widebar{x}^{(m)}_{r,k,c}}{\widehat{\sigma}_{r,k,t}}\right].
\end{aligned}
\end{equation*}
Here $\widehat{M}^{\SMD}_k$ reflects balancing of $X_k$ for the overall matched sample, with $|\widebar{x}^{(m)}_{k,t}-\widebar{x}^{(m)}_{k,c}|/\widehat{\sigma}^{(m)}_{k,t}$ being the standardized mean difference of $X_k$ in the overall matched sample; $\widehat{M}^{\SMD}_{r,k}$ reflects balancing of $X_k$ for the matched sample for subgroup $r$, with $|\widebar{x}^{(m)}_{r,k,t}-\widebar{x}^{(m)}_{r,k,c}|/\widehat{\sigma}^{(m)}_{r,k,t}$ being the standardized mean difference of $X_k$ in the matched sample for subgroup $r$.
The objective function is the sum of squares of these estimates, i.e.,
\begin{equation}
F^{\SMD}=\sum_{k=1}^K \left(\widehat{M}^{\SMD}_{k}\right)^2+\sum_{r=1}^R\sum_{k=1}^K \left(\widehat{M}^{\SMD}_{r,k}\right)^2.
\end{equation}

We consider nearest neighbour matching with caliper \citep*{rosenbaum1985constructing} to remove possible bias of matching. For each subgroup $r$, we set the caliper to one fourth of the standard deviation of the logit of estimated propensity scores for units in this subgroup; each treated unit in this subgroup is matched with a control unit in this subgroup whose logit of estimated propensity score is closest to and within the caliper of that of the former, and units that cannot be matched are dropped from further analysis. We allow replacement and ties in matching; hence the weights for matched treated units are all equal to one, but the weights for matched control units may be smaller or larger than one.


\subsection{Weighting-based Criterion}
The second criterion is related to covariate balance for propensity score weighting. For unit $i$ ($i=1,\cdots,N$), let $x_{ik}$ denote the value of the $X_k$. In the weighted sample, each treated unit is weighted by one and each control unit is weighted by $\hat{e}_i/(1-\hat{e}_i)$. Let
$\widebar{x}_{k,t}$ denote the mean of $x_{ik}$ for the treated units in the overall sample, and let $\widebar{x}_{r,k,t}$ denote the mean of $x_{ik}$ for the treated units in the sample for subgroup $r$.
The estimators of the moments in \eqref{eq:moment-weighting} are
\begin{equation*}
\begin{aligned}
\widehat{M}^{\PSW}_k&=\frac{1}{N}\left[N_t\widebar{x}_{k,t}-\sum_{Z_i=0}\frac{\hat{e}_i}{1-\hat{e}_i}x_{ik}\right],\\
\widehat{M}^{\PSW}_{(r)}&=\frac{1}{N}\left[N_{r,t}-\sum_{G_i=r,Z_i=0}\frac{\hat{e}_i}{1-\hat{e}_i}\right],\\
\widehat{M}^{\PSW}_{r,k}&=\frac{1}{N}\left[N_{r,t}\widebar{x}_{r,k,t}-\sum_{G_i=r,Z_i=0}\frac{\hat{e}_i}{1-\hat{e}_i}x_{ik}\right].
\end{aligned}
\end{equation*}
The objective function is the sum of squares of these estimates, i.e.,
\begin{equation}
F^{\PSW}=\sum_{k=1}^K \left(\widehat{M}^{\PSW}_{k}\right)^2+\sum_{r=1}^R \left(\widehat{M}^{\PSW}_{(r)}\right)^2+\sum_{r=1}^R\sum_{k=1}^K \left(\widehat{M}^{\PSW}_{r,k}\right)^2.
\end{equation}

The moment conditions for $M_k^{\PSW}$ and $M_{(r)}^{\PSW}$ have been used previously by the CBPS method \citep{imai2014covariate}, which uses the generalized method of moments (GMM) to incorporate these moment conditions and the score condition for the maximum likelihood in estimating the propensity score. The total number of moment conditions exceeds that of model parameters, thus giving a set of overidentifying restrictions.
The overidentifying restrictions generally improve asymptotic efficiency but may result in a poor finite sample performance, and hence \cite{imai2014covariate} used the `continuous updating' GMM estimator of \cite{Hansen1996GMM} which has better finite-sample properties than the usual optimal GMM estimator. This approach, however, is not appropriate for incorporating into GMM estimation the additional $RK$ moment conditions for $M_{r,k}^{\PSW}$. First, the `continuous updating' GMM method cannot converge in one week for any of our simulation in Section 4. Second, the estimate of $M_{r,k}^{\PSW}$ only involves the sample in subgroup $r$ (rather than the overall sample), imposing further difficulty for obtaining good finite sample performance.

\subsection{Estimating the SBPS}

For each subgroup, we choose between the overall sample fit and the subgroup sample fit. Specifically, for $r=1,\cdots,R$, let $S_r=1$ if the propensity scores for units in subgroup $r$ are estimated by fitting the model in \eqref{eq:psmodel1} using the overall sample, and let $S_r=2$ if these propensity scores are estimated by fitting the model in \eqref{eq:psmodel2} using the subgroup sample.  Let $\bm{S}=(S_1,\cdots,S_R)$. The criterion $F^{\SMD}$ or $F^{\PSW}$ can be regarded as a function of $\bm{S}$.

If $R$ is small, we can exhaustively search through all $2^R$ possible combinations for $\bm{S}$ to find the optimal combination that minimizes  $F^{\SMD}$ or $F^{\PSW}$. If $R$ is large, we propose to use the following stochastic search algorithm to find $\bm{S}$ that minimizes $F^{\SMD}$ or $F^{\PSW}$.
\begin{enumerate}
\item Initialize $\bm{S}_{\min}$ to be a vector of all ones, which corresponds to the traditional method that uses the overall sample fit to estimate propensity scores for all subgroups. Set $F^{\SMD}_{\min}$ or $F^{\PSW}_{\min}$ to be the corresponding value of $F^{\SMD}$ or $F^{\PSW}$.
\item For $l=1,\cdots,L$, repeat the following steps.
\begin{itemize}
\item For $r=1,\cdots,R$, randomly initialize $S_r=1$ or $S_r=2$.
\item Randomly permutate $\{1,\cdots,R\}$ to get $\{A_1,\cdots,A_R\}$.  Repeat the following step until there is no change in the elements of $\bm{S}$.
\begin{itemize} \item For $r=1,\cdots,R$, update $S_{A_r}$ to the value that gives a smaller value of $F^{\SMD}$ or $F^{\PSW}$ while fixing the other elements in $\bm{S}$, $\{S_{A_{r'}}, r'\neq r\}$.
\end{itemize}
\item If the value of $F^{\SMD}$ or $F^{\PSW}$ is smaller than $F^{\SMD}_{\min}$ or $F^{\PSW}_{\min}$, then set $F^{\SMD}_{\min}=F^{\SMD}$ or $F^{\PSW}_{\min}=F^{\PSW}$, and set $\bm{S}_{\min}=\bm{S}$.
\end{itemize}
\end{enumerate}

When $R$ is large, results of the stochastic search algorithm depend on the random initial values and the random orders of updates for $\bm{S}$.
 Although this algorithm cannot guarantee to find the globally optimal combination for $\bm{S}$, it can guarantee to find some locally optimal combination for $\bm{S}$ that gives a smaller value of $F^{\SMD}$ or $F^{\PSW}$ than the traditional method.

\section{A Simulation}\label{sec:simu1}


Consider a scenario with 20 subgroups and 100 sampled units in each subgroup, and hence $R=20$ and $N_r=100$ for $r=1,\cdots,R$. Assume that there are four covariates: $X_{1} \sim N(\mu_r,1)$ if $G=r$, where $\mu_r=-3+6(r-1)/(R-1)$ (i.e., $\mu_r$ varies from $-3$ to $3$ with the same increment $6/(R-1)$); $X_{2} \sim Unif(0,1)$; $X_{2} \sim N(0,1)$; $X_{4}\sim Bernoulli(0.4)$. The treatment assignment is generated using the following logistic model:
\begin{equation}
\mbox{logit}\left[e(\bm{X},G)\right]=\sum_{r=1}^R \delta_{r}\one\{G=r\} +\alpha_{1}X_{1}+\alpha_{2}X_{2}+\alpha_{3}X_{3}+\alpha_{4}X_{4}+\alpha_{5}X^2_{1}+\alpha_{6} X_{1}X_{4},
\end{equation}
where $\bm{\alpha}\equiv (\alpha_{1},\cdots,\alpha_{6})=(-1.5,-0.5,0.5,-0.5,0.5,0.5)$ and the fixed effects are $\delta_{r}=-1+2(r-1)/(R-1)$ (i.e. $\delta_r$ varies from $-1$ to $1$ with the same increment $2/(R-1)$). The outcome $Y$ is generated using the following linear model:
\begin{equation}
Y =\beta_{0}+\sum_{r=1}^R \eta_{r} [\one\{G=r\}Z]+\beta_{1}X_{1}+\beta_{2}X_{2}+\beta_{3}X_{3}+\beta_{4}X_{4}+\beta_{5}X^2_{1}+\beta_{6} X_{1}X_{4}+\varepsilon,
\end{equation}
where $(\beta _{0},\beta_{1},...,\beta_{6})=(200,20,10,10,10,-5,10)$, the true subgroup ATTs are given by $\tau_r=\eta_{r}=-10+20(r-1)/(R-1)$ (i.e. $\eta_{r}$ varies from $-10$ to $10$ with the same increment $20/(R-1)$), and the noise term $\varepsilon\sim N(0,1)$.

We compare results from the traditional method, the CBPS method, the SBPS method that optimizes $F^{\SMD}$ (SBPS-SMD hereafter) and the SBPS method that optimizes $F^{\PSW}$ (SBPS-PSW hereafter).  For the traditional method and the CBPS method, the propensity score model in \eqref{eq:psmodel1} is fitted to the overall sample to estimate $\hat{e}_i$ for all units, where $\bm{X}=(X_{1},X_{2},X_{3},X_{4},X_{1}^2,X_{1}X_{4})^{\top}$ if the propensity score model is correctly specified, and $\bm{X}=(X_{1},X_{2},X_{3},X_{4})^{\top}$ if the propensity score model is misspecified.  For the SBPS-SMD  and SBPS-PSW methods, to estimate propensity scores $\hat{e}_i$ for units in subgroup $r$, the propensity score model in \eqref{eq:psmodel1} is fitted to the overall sample if $S_r=1$, and the propensity score model in \eqref{eq:psmodel2} is fitted to the subgroup sample if $S_r=2$; we set the number of iterations in the stochastic search algorithm to be $L=1,000$. Once $\hat{e}_i$ is obtained using a given method, propensity score matching or weighting can be used to estimate $\tau_r$.

We generate $V=1000$ data sets. For the $v$'th data set, let $\hat{\tau}_{r,v}$ denote the value of $\hat{\tau}_r$ ($\hat{\tau}_r^{Dir}$ or $\hat{\tau}_r^{PSW}$). To evaluate the various methods, we consider three performance measures. The first measure is absolute bias,
\begin{equation}\label{eq:abias}
B_r=\left|\frac{1}{V}\sum_{v=1}^V \hat{\tau}_{r,v}-\tau_r\right|,
\end{equation}
and the second measure is root mean squared error (RMSE),
\begin{equation}\label{eq:abias}
E_r=\sqrt{\frac{1}{V}\sum_{v=1}^V (\hat{\tau}_{r,v}-\tau_r)^2}.
\end{equation}
We also consider 95\% confidence intervals of the form  $\hat{\tau}_{r,v}\pm 1.96 \times \widehat{SE}(\hat{\tau}_{r,v})$, where $\widehat{SE}(\hat{\tau}_{r,v})$ is the standard error of $\hat{\tau}_{r,v}$ estimated using 1,000 bootstrap samples. The third performance measure $C_r$ is the proportion of confidence intervals that cover $\tau_r$. We then average the performance measures across subgroups to get
\begin{equation*}
\widebar{B}=\frac{1}{R}\sum_{r=1}^R B_r,\quad
\widebar{E}=\frac{1}{R}\sum_{r=1}^R E_r\quad\mbox{and}\quad
\widebar{C}=\frac{1}{R}\sum_{r=1}^R C_r.
\end{equation*}

\begin{figure}[t]
  \centering
  \includegraphics[width=7in]{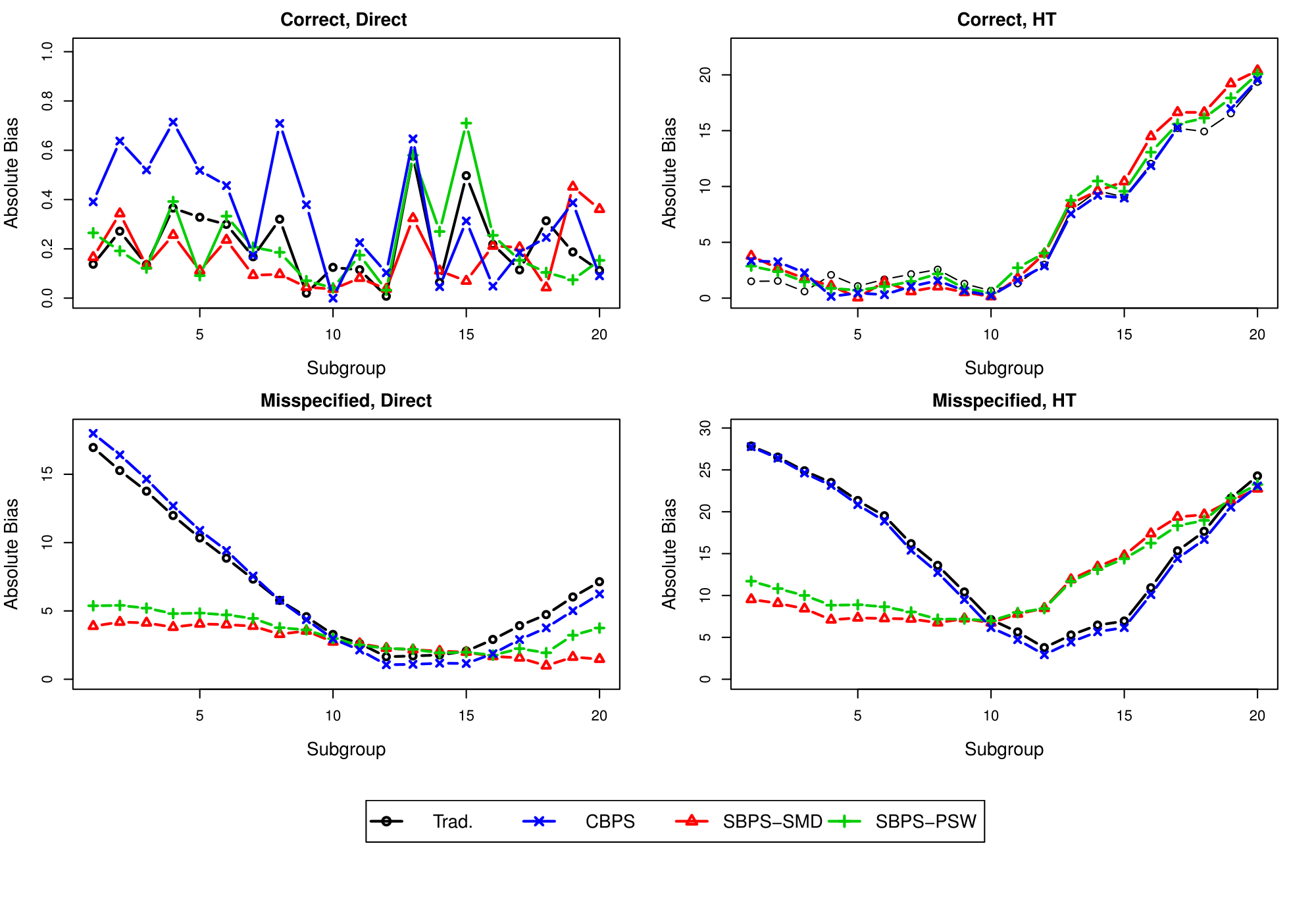}
  \caption{Absolute Bias $B_r$ for Each Subgroup for the Four Propensity Score Estimation Methods (Simulation)}\label{fig:bias-20}
\end{figure}

\begin{figure}[t]
  \centering
  \includegraphics[width=7in]{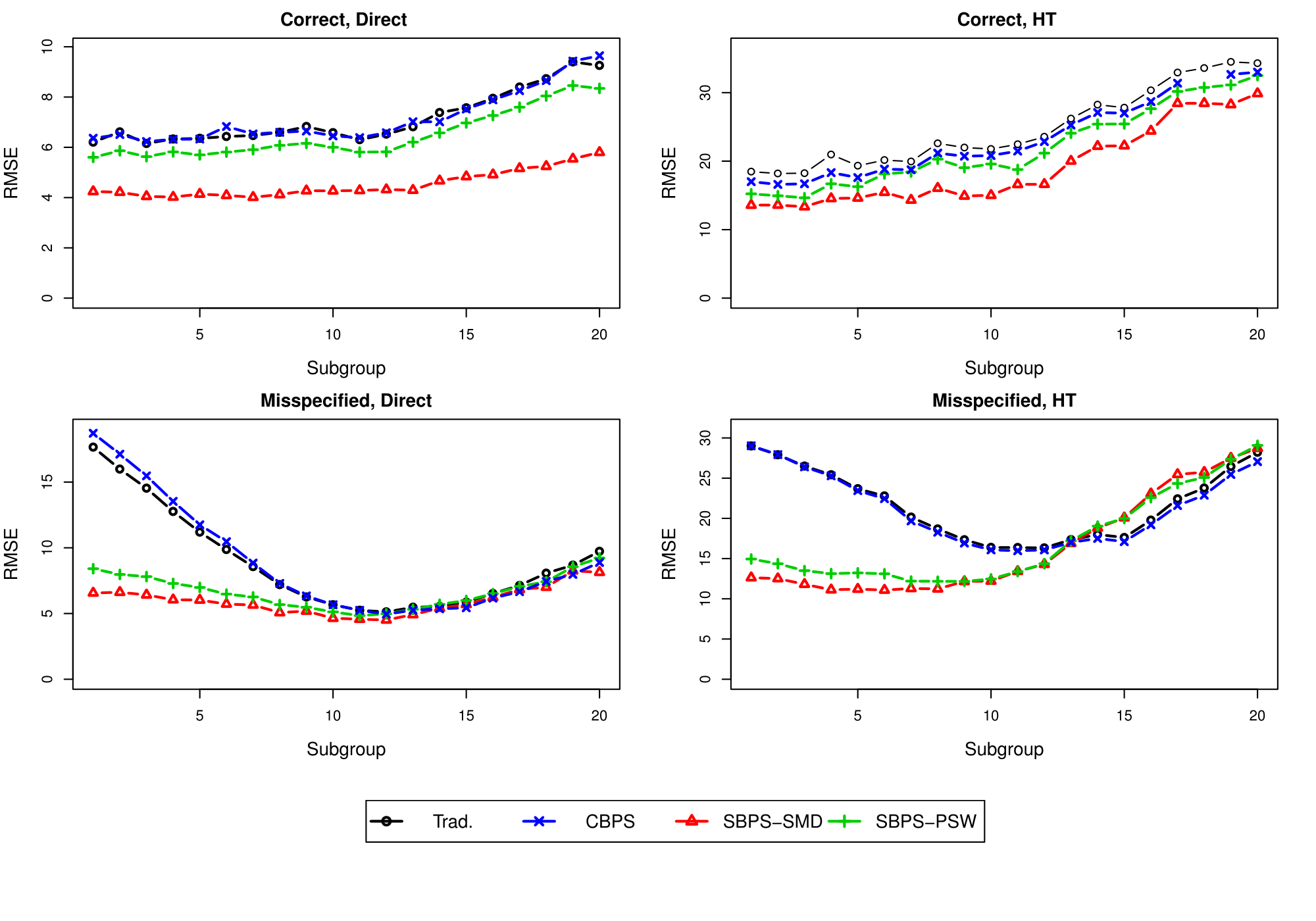}\\
  \caption{RMSE $E_r$ for Each Subgroup for the Four Propensity Score Estimation Methods (Simulation)}\label{fig:rmse-20}
\end{figure}

\begin{figure}[t]
  \centering
  \includegraphics[width=7in]{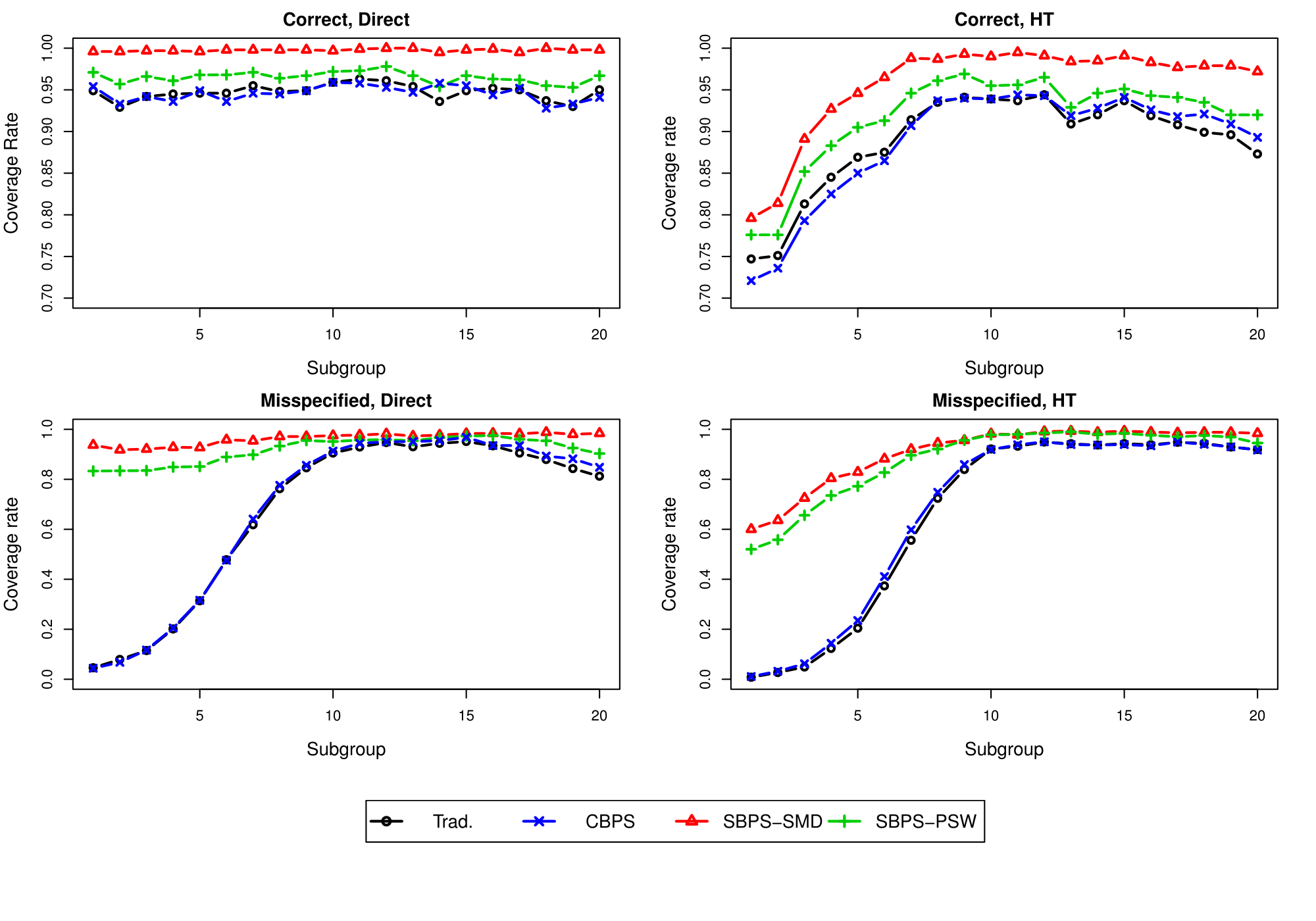}\\
  \caption{Coverage Rate $C_r$ for Each Subgroup for the Four Propensity Score Estimation Methods  (Simulation)}\label{fig:coverage-20}
\end{figure}

When the propensity score model is correctly specified or misspecified, and when $\hat{\tau}_r^{Dir}$ or $\hat{\tau}_r^{PSW}$ is used to estimate the subgroup treatment effect $\tau_r$, Figures \ref{fig:bias-20}-\ref{fig:coverage-20} plot the values of $B_r$, $E_r$ and $C_r$ ($r=1,\cdots,R$) for the four propensity score estimation methods, and Table \ref{table:results-20} reports the values of $\widebar{B}$, $\widebar{E}$ and $\widebar{C}$ for the four propensity score estimation methods. Regardless of whether the propensity score model is correctly specified or misspecified, and regardless of the propensity score estimation method, the PSW estimator has larger absolute bias and larger RMSE than the direct matching estimator.  This can be attributed to the fact that in this simulation the propensity score does not have common support in the treated and control groups, which is taken into account by the direct matching estimator but not by the PSW estimator. We therefore focus on the direct matching estimator. When the model is correctly specified, SBPS-SMD has a smaller average RMSE than the traditional method or the CBPS method (4.53 versus 7.15 or 7.16).  When the model is misspecified, compared to the traditional method or the CPBS method, SBPS-SMD has a smaller average absolute bias (2.80 versus 6.63 or 6.46), a smaller average RMSE (5.98 versus 8.85 or 8.93), and also considerably improves the coverage rate (0.96 versus 0.67 or 0.68).  SBPS-PSW  does not perform as well as SBPS-SMD. For most simulated data sets, the performance of the CBPS method is similar to that of the traditional method, but when combined with the PSW estimation, the CBPS method may yield extremely large errors for some data sets when the model is correctly specified, hence giving the extremely large values of bias (246.05) and RMSE (7983.40) in Table \ref{table:results-20}. We recommend to use SBPS-SMD combined with the direct matching estimator.

For SBPS-SMD, the average proportion of subgroups using subgroup sample fit (i.e., with $S_r=2$) is 73\% when the propensity score model is correctly specified, and is 87\% when the propensity score is incorrectly specified. Hence in order to achieve better subgroup covariate balance, the subgroup sample fit instead of the overall sample fit is used for a fairly large proportion of subgroups.

We also conducted the simulation with $R=40$ and $N_r=50$ for $r=1,\cdots,R$ to represent the case with larger number of smaller subgroups, and the results are similar.

\renewcommand{\baselinestretch}{1}

\begin{table}[!htb]
\renewcommand\arraystretch{1.2}
\centering
\begin{tabular}{cl|cccc|cccc} \hline\hline

&&\multicolumn{4}{c|}{Direct}&\multicolumn{4}{c}{PSW}\\
\cline{3-10}
&&\multirow{2}{*}{Traditional}&\multirow{2}{*}{CBPS}&SBPS&SBPS&\multirow{2}{*}{Traditional}&\multirow{2}{*}{CBPS}&SBPS&SBPS \\
&&&&-SMD&-PSW&&&-SMD&-PSW\\ \hline
\multirow{3}{*}{Correct}&$\widebar{B}$&0.09&0.28&0.01&0.09&5.05&246.05&6.32&5.92\\
&$\widebar{E}$&7.15&7.16&4.53&6.48&24.78&7983.40&19.13&22.02\\
&$\widebar{C}$&0.95&0.95&1.00&0.97&0.89&0.89&0.96&0.92\\ \hline
\multirow{3}{*}{Misspecified}&$\widebar{B}$&6.63&6.46&2.80&3.45&15.44&14.71&11.68&12.12\\
&$\widebar{E}$&8.85&8.93&5.98&6.66&21.71&21.27&16.56&17.19\\
&$\widebar{C}$&0.67&0.68&0.96&0.92&0.66&0.67&0.91&0.88\\
\hline\hline
\end{tabular}
\caption{Average Performance Measures for the Four Propensity Score Estimation Methods  (Simulation)}\label{table:results-20}
\end{table}

\renewcommand{\baselinestretch}{1.75}

\section{Application to the SHIW Data}
\subsection{Real Data Analysis}
We apply the proposed method to the 1993-1995 Italy Survey of Household Income and Wealth (SHIW) data to evaluate the effect of having debit cards on household spending. Here the treatment variable equals one if the household possesses one and only one debit card and zero if the household does not possess debit cards during year 1993-1995, and the outcome is the average monthly household consumption on all consumer goods in 1995. The covariates, listed in Table \ref{table:SHIWvars}, include the lagged outcome in 1993, background demographic and social variables referred either to the household or to the head householder, number of banks and yearly-based average interest rate in the province where the household lives.

Based on the SHIW data, \cite{mercatanti2014debit} found statistically significant positive average effects of possessing debit cards on household spending for the whole population. The findings can be explained by the mental accounting theory which describes how consumers and their households organize, evaluate and record their economic activities \citep{thaler1985mental,thaler1990anomalies,thaler1999mental}. Non-cash payment instruments, such as debit and credit cards, decouple purchases from payment and mix different purchases. Consumers have cognitive biases and hence may spend more when they use the non-cash methods.

In this paper we further investigate the subgroup effects of debit cards possession on consumption for households with different income levels. Because the marginal propensity to consume \citep{keynes1936general}, or the additional amount of consumption induced by a unit of additional income, decreases with income, consumers' additional amount of consumption using non-cash methods may also decrease with income. We partition households in the SHIW data into six income groups based on the overall household income in units of thousands of Italian liras (with $G$ correspondingly equal to 1 to 6): $\leq20,000$, $20,000-30,000$, $30,000-40,000$, $40,000-50,000$, $50,000-60,000$, and $>60,000$, and investigate the effects of debit cards possession on the monthly household consumption for each income group. Table \ref{table:SHIWnum} presents the number of treated and control units in each income group.

\renewcommand{\baselinestretch}{1}

\begin{table}[htbp]
\begin{center}
\begin{tabular}{c|l} \hline\hline
Variable&Description\\ \hline
$Y$& Average monthly consumption on all consumer goods in 1995\\
& (in thousands of Italian liras)\\ \hline
$Z$&=1 if the household possesses one and only one debit card\\
&=0 if the household does not possess debit cards\\ \hline
$G$&Income group based on the overall household income (in thousands of Italian Liras)\\
&=1 if $\leq20,000$\\
&=2 if $20,000-30,000$\\
&=3 if $30,000-40,000$\\
&=4 if $40,000-50,000$\\
&=5 if $50,000-60,000$\\
&=6 if $>60,000$\\ \hline
$X_1$&Average monthly consumption on all consumer goods in 1993\\
$X_2$&The overall household wealth\\
$X_3$&The Italian geographical macro-area where the household lives:\\
&north (baseline), center, south and islands.\\
$X_4$&The number of inhabitants of the town where the household lives:\\
&$<20,000$ (baseline), 20,000-40,000, 40,000-500,000, $>500,000$\\
$X_5$&The number of household members:\\
&1 (baseline), 2, 3, 4, $>4$\\
$X_6$&The number of earners in the household:\\
&1 (baseline), 2, 3, $>3$\\
$X_7$&Average age of the household:\\
&$<31$ (baseline), 31-40, 41-50, 51-65, $>65$\\
$X_8$&Education of the head of the household:\\
&None (baseline), elementary school, middle school, high school, university\\
$X_9$&Age of the head of the household: \\
&$<31$ (baseline), 31-40, 41-50, 51-65, $>65$\\
$X_{10}$&Number of banks\\
$X_{11}$&Average interest rate in the province where the household lives \\
\hline \hline
\end{tabular}
\caption{Variables in The SHIW Data}\label{table:SHIWvars}
\end{center}
\end{table}

\renewcommand{\baselinestretch}{1}
\begin{table}[htbp]\footnotesize
\begin{tabular}{l|cccccc} \hline\hline
&\multicolumn{6}{c}{Income Group}\\
&$\leq20,000$&$20,000-30,000$&$30,000-40,000$&$40,000-50,000$&$50,000-60,000$&$>60,000$\\ \hline
\#Treated&17&47&41&36&26&49\\
\#Control&126&189&168&119&77&105\\
\hline \hline
\end{tabular}
\caption{Number of Treated and Control Units in Each Income Group in The SHIW Data}\label{table:SHIWnum}
\end{table}

We applied the four propensity score estimation methods to the data, where for the SBPS-SMD and SBPS-PSW methods exhaustive search is used to find the optimal combination for $\bm{S}$. Table \ref{table:SHIWeffects} show the estimated effects and the associated $p$-values.  Here the $p$-value is calculated as $2\Phi^{-1}\left(\left|\hat{\tau}_r/\widehat{SE}(\hat{\tau}_{r})\right|\right)$, where $\Phi$ is the distribution function for a standard normal distribution, and $\widehat{SE}(\hat{\tau}_{r})$ is the standard error of $\hat{\tau}_{r}$ estimated using 1,000 bootstrap samples. We consider a significance level of 0.05. For the direct matching estimate, only the SBPS-SMD method found a significant positive effect 681.25 for the lowest income group, which is consistent with our conjecture that the effects of debit cards possession on household spending decrease with income.  For the PSW estimate, all four methods found significant positive effects (323.78, 312.98, 528.41 and 323.78) for the lowest income group, but the traditional method and the SBPS-SMD method also found a significant positive effect 367.34 for the highest income group, which seems inconsistent with our conjecture.

In order to account for simultaneous estimation of effects for multiple subgroups, which is known as the multiple testing issue, we use the false discovery rate method by \cite{benjamini1995controlling} to calculate the adjusted $p$-values in Table \ref{table:SHIWeffects}. This method controls for the false discovery rate, or the expected proportion of rejected null hypotheses that are incorrect rejections. It is more powerful than methods based on the family-wise error rate, such as the Bonferroni correction. Judging from the adjusted $p$-value, only the SBPS-SMD method yields significantly positive effect for the lowest income group, regardless of whether the direct matching estimate or the PSW estimate is used. This is consistent with our previous conjecture.

\renewcommand{\baselinestretch}{1}

\begin{table}
\centering
\begin{tabular}{ll|rcc|rcc} \hline\hline
\multirow{2}{*}{Method}&\multirow{2}{*}{Group}&\multicolumn{3}{|c|}{Direct}&\multicolumn{3}{c}{PSW}\\ \cline{3-8}
&&Effect&$p$-value&Adj. $p$&Effect&$p$-value&Adj. $p$\\ \hline
\multirow{6}{*}{Traditional}&$\leq20,000$&273.53&0.15&0.44&323.78&0.04&0.13\\
&$20,000-30,000$&-105.75&0.35&0.53&-8.38&0.93&0.93\\
&$30,000-40,000$&-246.46&0.21&0.44&-133.32&0.23&0.45\\
&$40,000-50,000$&-71.43&0.69&0.69&-35.24&0.78&0.93\\
&$50,000-60,000$&134.62&0.57&0.69&133.87&0.42&0.63\\
&$>60,000$&311.22&0.22&0.44&367.34&0.04&0.13\\ \hline
\multirow{6}{*}{CBPS}&$\leq20,000$&255.88&0.25&0.50&312.93&0.05&0.17\\
&$20,000-30,000$&51.74&0.64&0.96&-13.53&0.88&0.88\\
&$30,000-40,000$&-203.98&0.24&0.50&-139.55&0.20&0.41\\
&$40,000-50,000$&14.29&0.93&0.96&-56.62&0.66&0.80\\
&$50,000-60,000$&-12.00&0.96&0.96&125.66&0.45&0.68\\
&$>60,000$&375.00&0.17&0.50&349.39&0.06&0.17\\ \hline
\multirow{6}{*}{SBPS-SMD}&$\leq20,000$&681.25&0.01&{\bf 0.03}&528.41&0.01&{\bf 0.03}\\
&$20,000-30,000$&-139.57&0.20&0.44&-29.98&0.78&0.78\\
&$30,000-40,000$&60.57&0.72&0.72&-49.09&0.67&0.78\\
&$40,000-50,000$&-71.43&0.69&0.72&-35.24&0.78&0.78\\
&$50,000-60,000$&125.00&0.58&0.72&264.30&0.17&0.34\\
&$>60,000$&311.22&0.22&0.44&367.34&0.04&0.13\\ \hline
\multirow{6}{*}{SBPS-PSW}&$\leq20,000$&273.53&0.15&0.60&323.78&0.04&0.23\\
&$20,000-30,000$&-139.57&0.20&0.60&-29.98&0.78&0.78\\
&$30,000-40,000$&60.57&0.72&0.72&-49.09&0.67&0.78\\
&$40,000-50,000$&-71.43&0.69&0.72&-35.24&0.78&0.78\\
&$50,000-60,000$&134.63&0.57&0.72&133.87&0.42&0.78\\
&$>60,000$&193.62&0.47&0.72&283.66&0.14&0.42\\
\hline \hline
\end{tabular}
\caption{Estimated Treatment Effects for Each Income Group in the SHIW Data Using the Four Propensity Score Estimation Methods, and The Associated p-values and Adjusted p-values}\label{table:SHIWeffects}
\end{table}
\renewcommand{\baselinestretch}{1.75}

\subsection{SHIW-based Simulations}

We further conduct a simulation based on the SHIW data to better understand the performance of the methods. We fix the covariates $\bm{X}$ and the group label $G$ as in the SHIW data. We generate the treatment assignment using the logistic model \eqref{eq:psmodel1}, where $\bm{\alpha}$ is set to the estimated coefficients of $\bm{X}$ from fitting \eqref{eq:psmodel1} to the real data, and $(\delta_1,\cdots,\delta_6)=(0.3,0.4,0.5,0.6,0.7,0.8)$. The outcome $Y$ is generated using the following linear model:
\begin{equation}\label{eq:real-Y}
Y =\beta_{0}+\sum_{r=1}^R \eta_{r} [\one\{G=r\}Z]+\bm{\beta}^{\top}\bm{X}+\varepsilon.
\end{equation}
We conduct propensity score matching over the real data based on the estimated propensity score from fitting \eqref{eq:psmodel1} to the real data, apply \eqref{eq:real-Y} to the matched sample, and set $\beta_0$ and $\bm{\beta}$ to be the estimated intercept and coefficients of $\bm{X}$. We then set $(\eta_1,...,\eta_{6})=(500,400,300,200,100,0)$ and generate $\varepsilon \sim N(0,50^2)$.

We estimate the propensity score using the traditional method, the CBPS method, the SBPS-SMD and SBPS-PSW methods similarly as in Section \ref{sec:simu1}, except that for the SBPS-SMD and SBPS-PSW methods exhaustive search is used to find the optimal combination for $\bm{S}$. When the propensity score model is correctly specified, we include all covariates ($X_1$ to $X_{11}$) in Table \ref{table:SHIWvars}; when the propensity score model is misspecified, we includes all covariate other than $X_2$ (the overall household wealth).

Table \ref{results-realsimu} reports the values of $\widebar{B}$, $\widebar{E}$ and $\widebar{C}$ for the four propensity score estimation methods. In this simulation, the propensity score has better overlap in the treated and control groups than in the simulation in Section 4, and the PSW estimator gives smaller average RMSE than the direct matching estimator. When the model is correctly specified, for the direct matching estimator, the SBPS-SMD method has smaller average RMSE than the traditional method or the CBPS method; when the model is misspecified, for both the direct matching estimator and the PSW estimator, the SBPS-SMD method has smaller average absolute bias and smaller average RMSE than the traditional method or the CBPS methods.  The SBPS-PSW method again does not perform as well as the SBPS-SMD method. In terms of coverage, here the SBPS-SMD method does not show any advantage over the traditional or the CBPS method.

\renewcommand{\baselinestretch}{1}

\begin{table}[!htb]
\renewcommand\arraystretch{1.2}
\centering
\begin{tabular}{cl|cccc|cccc} \hline\hline

&&\multicolumn{4}{c|}{Direct}&\multicolumn{4}{c}{PSW}\\
\cline{3-10}
&&\multirow{2}{*}{Traditional}&\multirow{2}{*}{CBPS}&SBPS&SBPS&\multirow{2}{*}{Traditional}&\multirow{2}{*}{CBPS}&SBPS&SBPS \\
&&&&-SMD&-PSW&&&-SMD&-PSW\\ \hline
\multirow{3}{*}{Correct}&$\widebar{B}$&5.93&6.59&9.74&3.30&15.47&15.73&19.09&10.56\\
&$\widebar{E}$&112.69&112.62&104.08&109.57&90.01&87.29&89.68&86.66\\
&$\widebar{C}$&0.95&0.95&0.98&0.97&0.95&0.96&0.98&0.97\\ \hline
\multirow{3}{*}{Misspecified}&$\widebar{B}$&40.29&39.70&30.25&45.59&41.98&43.21&26.33&46.70\\
&$\widebar{E}$&124.00&122.17&114.09&122.30&103.86&102.58&97.90&104.59\\
&$\widebar{C}$&0.94&0.95&0.97&0.96&0.94&0.94&0.97&0.96\\

\hline\hline
\end{tabular}
\caption{Average Performance Measures for the Four Propensity Score Estimation Methods (SHIW-based simulation)}\label{results-realsimu}
\end{table}

\section{Discussion}

In this paper, we have focused on the case when the subgroups are defined by one covariate, and have demonstrated that the SBPS-SMD method combined with direct matching estimator has the best overall performance in estimating subgroup treatment effects. In a more general situation, the subgroups can be defined by multiple covariates, e.g., education and income. We now discuss how the SBPS method can be extended to such cases.

Let $G_h$ ($h=1,\cdots,H$) denote the labels for covariates that define the subgroups, and suppose that $G_h$ can take values among $1,\cdots,R_h$. Hence the total number of subgroups is $\prod_{h=1}^H R_h$. Let $\bm{X}$ now denote the covariates other than those defining the subgroups. The true propensity score is denoted by $e(\bm{X},G_1,\cdots,G_H)$.

Let $\{h_1,\cdots,h_j\}$ denote any $j$-subset (i.e., a subset with $j$ elements) of $\{1,\cdots,H\}$ ($j\in\{1,\cdots,H\}$). It is easy to extend Result 1 to show that the true propensity score satisfies
\begin{equation*}
\bm{X}\perp Z|\{G_{h_1}=r_{h_1},\cdots,G_{h_j}=r_{h_j},e(\bm{X},G_1,\cdots,G_H)\}
\end{equation*}
for any $j$-subset and any values $r_{h_1}\in\{1,\cdots,R_{h_1}\},\ \cdots\ ,r_{h_j}\in\{1,\cdots,R_{h_j}\}$ ($j\in\{1,\cdots,H\}$). Hence besides balancing the distribution of $\bm{X}$ and $(G_1,\cdots,G_H)$ for the overall population, the true propensity score can also balance the distribution of $\bm{X}$ within each subgroup defined by any $j$-subset ($j\in\{1,\cdots,H\}$).

For example, if the subgroups are defined by education and income, then the true propensity score can balance: (a) the distribution of $\bm{X}$; (b) the distribution of education; (c) the distribution of income; (d) the distribution of $\bm{X}$ within each education level; (e) the distribution of $\bm{X}$ within each income level; (f) the distribution of $\bm{X}$ within each subgroup cross-classified by education and income. We will use this example further to illustrate how the SBPS method can be extended. For each (lowest-level) subgroup cross-classified by education and income, the propensity scores for units in this subgroup can be estimated using the overall sample, the sample for the corresponding education level (which includes units with the same education level but different income levels), the sample for the corresponding income level (which includes units with the same income level but different education levels), or the subgroup-specific sample. The combination of estimation samples for all subgroups can be chosen in order to optimize an objective function $F^{\SMD}$ or $F^{\PSW}$ that accounts for the set of covariate-balancing moment conditions for all of (a)-(f) mentioned above. The SBPS can again be estimated by a stochastic search algorithm.

\bibliographystyle{jasa3}
\bibliography{sbpsbibfile}







\end{document}